\documentclass[pre,oneocolumn,superscriptaddress,nofootinbib]{revtex4}

\usepackage{graphicx}
\usepackage{dcolumn}

\usepackage{eucal}
\usepackage[dvips]{epsfig}
\usepackage{amssymb}

\usepackage{amsmath,amsthm}

\newcommand{\be}{\begin{eqnarray}}
\newcommand{\ee}{\end{eqnarray}}
\newcommand{\bse}{\begin{subequations}}
\newcommand{\ese}{\end{subequations}}

\newcommand{\bnum}{\begin{enumerate}}
\newcommand{\enum}{\end{enumerate}}

\newcommand{\bit}{\begin{itemize}}
\newcommand{\eit}{\end{itemize}}

\newcommand{\bc}{\begin{cases}}
\newcommand{\ec}{\end{cases}}

\newcommand{\bpm}{\begin{pmatrix}}
\newcommand{\epm}{\end{pmatrix}}

\newcommand{\bvm}{\begin{vmatrix}}
\newcommand{\evm}{\end{vmatrix}}

\newcommand{\mrm}{\mathrm}

\newcommand{\gd}{\delta}
\newcommand{\eps}{\epsilon}

\newcommand{\go}{\omega}

\newcommand{\gr}{\rho}

\newcommand{\Go}{\Omega}

\newcommand{\p}{\partial}

\newcommand{\IDoS}{\Omega}
\newcommand{\DoS}{\omega}

\newcommand{\Ent}{S}
\newcommand{\Temp}{T}

\newcommand{\EntG}{\Ent_{\mathrm{G}}}
\newcommand{\EntB}{\Ent_{\mathrm{B}}}

\newcommand{\TempG}{\Temp_{\mathrm{G}}}
\newcommand{\TempB}{\Temp_{\mathrm{B}}}

\newcommand{\Sch}{Schneider \textit{et al.}}
\begin{document}
\title{Reply to Schneider \textit{et al.}}

\author{J\"orn Dunkel}
\affiliation{Department of Mathematics, Massachusetts Institute of Technology, 
77 Massachusetts Avenue E17-412, 
Cambridge, MA 02139-4307, USA }

\author{Stefan Hilbert}
\affiliation{Exzellenzcluster Universe, Boltzmannstr. 2, D-85748 Garching, Germany}

\date{\today}
\begin{abstract} 
We restrict our response to pointing out the main physical, logical and historical errors in the Comment of \Sch~\cite{2014Schneider_Comment}. 
The claims by \Sch{} contradict exact results obtained by Gibbs~\cite{Gibbs}, Hertz~\cite{1910Hertz_1,1910Hertz_2} and Khinchin~\cite{Khinchin}, rendering the criticism in Ref.~\cite{2014Schneider_Comment} invalid.
\end{abstract}

\pacs{ }
\maketitle

In their Comment, \Sch{} claim that the canonical ensemble with negative parameter $\beta$ provides a consistent thermodynamic description for systems with bounded spectrum, as realized in their recent experiments~\cite{2013Braun}. This claim is incorrect for several reasons. As already discussed in our paper~\cite{2014DuHi_NatPhys},  Ref.~\cite{2013Braun} purports to have realized a stable population-inverted state\footnote{See p.55 in Ref.~\cite{2013Braun}, where it is stated that \textit{\lq\lq In contrast to metastable excited states (21), this isolated negative temperature ensemble is intrinsically stable and cannot decay into states of lower energy.\rq\rq} }, which can only be achieved in a thermally isolated system thus demanding a microcanonical description. As also discussed in our paper~\cite{2014DuHi_NatPhys}, ensemble equivalence does not hold for population-inverted systems, in contrast to the claims by \Sch{}. Furthermore, as explained in detail below, the spurious symmetries of entropy $S$ and energy $E$ advocated by \Sch{}  violate essential principles of thermodynamics, contradict experimental evidence and ignore generic stability requirements that any physical system has to satisfy. As a result,  \Sch{} are led to paradoxical conclusions~\cite{2013Braun,2010Rapp}, such as Carnot efficiencies $>1$~\cite{2013Braun,2010Rapp} which, according to standard thermodynamic theory, imply the possibility of \textit{perpetual motion}.

\par
As a key part of their argument, \Sch{} invoke a hypothetical symmetry $S(E) \to S(-E)$ of the entropy, motivated by recourse to a symmetry $E \to - E$ in the spectra of Ising-like model Hamiltonians. However,  the fact that certain artificially truncated Hamiltonians possess symmetric spectra does \emph{not} imply that low-energy states and high-energy states are physically or thermodynamically equivalent. Within conventional thermodynamics the internal energy is a \emph{directed} (ordered) quantity:  The assumption that a higher energy state is, in principle, physically distinguishable from a lower energy state is a basic axiom of thermodynamics -- without this empirically supported fact it would not be possible to distinguish whether a thermodynamic cycle produces work or absorbs work. \Sch{} ignore this defining aspect of thermodynamics by postulating that  $S(E)=S(-E)$ if the Hamiltonian is bounded and symmetric. If this postulate was reasonable, then the groundstate and the highest energy state would be thermodynamically equivalent. But even for symmetric Hamiltonians this is clearly not the case: A system occupying the groundstate can be easily heated (e.g., by injecting a photon) whereas it is impossible to add a photon to a system in the highest energy state. Thermodynamic entropy has to reflect this asymmetry -- otherwise the resulting theoretical framework has limited predictive power for real-world thermodynamic processes (see SI of Ref.~\cite{2014DuHi_NatPhys}).

\par
\Sch{} motivate their deficient entropy symmetry postulate by considering a hypothetical Hamiltonian $H'=-H$.  Although $H'$ is mathematically well-defined for artificially truncated spin Hamiltonians $H$, the reflected Hamiltonian $H'$ becomes physically ill-defined  if one considers the underlying untruncated (\lq full\rq)~Hamiltonian $H_\mrm{phys}$ instead. To our knowledge, all currently available experimental evidence suggests that all physically acceptable Hamiltonians $H_\mrm{phys}$ describing systems in our universe must be bounded from below (otherwise, we would not exist) whereas, to date, no evidence for strict upper energy bounds has been reported. That is, for any real Hamiltonian $H=H_\text{phys}$,  $H'=-H$ becomes unbounded from below, hence, belonging to the class of unstable Hamiltonians.  Thus, in essence, the arguments of \Sch{} rely on ill-conceived symmetry postulates that are not only in conflict with standard thermodynamics but also violate general stability requirements.  

\par
In this  context, we would like to stress again that a canonical description is generally incompatible with the experimental conditions described in Ref.~\cite{2013Braun}. A canonical ensemble approach is feasible when an energetically isolated system can be cleanly divided into a (sub)system of interest and a quasi-infinite bath. One of the main assumptions  underlying this division is that the bath is not affected when work is performed on the subsystem. However, according to the description of the experimental setup in Ref.~\cite{2013Braun}, the ultra-cold gas was kept in isolation and all atoms in the system were subject to the applied changes in the potential parameters.  Under these conditions,  the whole gas must be treated microcanonically as one isolated thermodynamic system. Moreover, if the density of states (DoS)  is a non-monotonic function of energy, ensemble equivalence is violated and the canonical description is not equivalent to the conceptually preferable microcanonical description.

\vspace{0.1cm}
\par
For completeness, we still address the main errors in the arguments listed under (i)-(vi) in the Comment.

\vspace{0.2cm}
\par
(i)  \Sch{} claim that \textit{ \lq\lq The entropy $\EntG$ is unphysical in the sense that it cannot be computed from the density matrix
alone. It furthermore depends on states that are energetically inaccessible to the system.\rq\rq}{}
This statement is incorrect because the entropy $\EntG=\ln \Go$ can be computed by simply integrating the normalization constant $\go$ of the microcanonical density operator, $\gr=\gd(E-H)/\go(E)$, over the energy to obtain the integrated DoS~$\Go$. Knowledge of $\Go$ is equivalent to knowledge $\go$. It is simply a matter of personal preference whether one considers cumulative distribution functions ($\Omega$) or densities ($\go$) as primary objects. The exact results summarized in Ref.~\cite{2014HiHaDu} further prove that only $\EntG$ satisfies the zeroth, first and second law simultaneously for the majority of physical systems. The fact that $\Omega$ accounts for states with energy $<E$ is key to proving~\cite{2014HiHaDu} that $\EntG$ generally satisfies Planck's second law (in contrast to the Boltzmann entropy $\EntB=\ln \eps\go$ adopted in Ref.~\cite{2013Braun}).

\vspace{0.2cm}
\par
(ii) \Sch{} claim that \textit{\lq\lq $\EntG$ cannot be connected to foundational concepts of modern statistical physics based on
information theory.\rq\rq{}} This statement is meaningless as it remains unclear what exactly they mean by \textit{\lq\lq foundational concepts of modern statistical physics based on information theory\rq\rq{}}. It seems advisable to focus on arguments that are, at least in principle,  falsifiable. Thermodynamics is a physical theory that aims to describe physical systems, which are governed by certain microscopic laws (Hamilton equations, Schr\"odinger equations, etc.), in terms of conserved charges and symmetry-breaking parameters~\cite{Callen}. A well-posed question is whether or not a certain thermostatistical formalism complies with both the microscopic dynamics and the laws of thermodynamics. This was the motivation for the work of Gibbs~\cite{Gibbs} and Hertz~\cite{1910Hertz_1,1910Hertz_2}, who both found that $\EntG$ achieves exactly this for isolated systems described by the microcanonical ensemble. In some cases, thermodynamic entropies can be identified with popular information measures, and  it may be theoretically pleasing when this happens, but there is no conceptual necessity for such an identification. Moreover, one could simply add $\EntG$ to the long list of information measures~\cite{1960Renyi,1978Wehrl}, which merely corroborates that item (ii) in Ref.~\cite{2014Schneider_Comment} is devoid of substance.

\vspace{0.2cm}
\par
(iii) \Sch{} state that \textit{\lq\lq In inverted situations the thermodynamic limit of the approach of Dunkel and Hilbert is
completely ill defined\rq\rq}, citing Ref.~\cite{2014VilarRubi} in this context. This as well as other statements in Ref.~\cite{2014VilarRubi} are incorrect  [see also item (v) below]. The thermodynamic limit (TDL)  of the Gibbs entropy~$\EntG$ is certainly well-defined, it just so happens that the temperature, which is merely one of the many secondary state variables in a microcanonical system diverges in certain energy ranges for certain systems. In the context of inverted systems, this divergence just signals that, on  macroscopic scales, population inversion cannot be created by ordinary thermal heating. At the same time, it should be kept in mind that TDLs are artificial mathematical constructs that is useful in the analysis of phase transitions -- but any real system is finite and, therefore, characterized by a finite positive Gibbs temperature. 

\vspace{0.2cm}
\par
(iv) \Sch{} state that \textit{\lq\lq $\EntG$ violates the second law of thermodynamics in the formulation that entropy cannot
decrease in an isolated system\rq\rq}. This statement is false in several regards, as already discussed in the Supplementary Information of our paper~\cite{2014DuHi_NatPhys}: The thermodynamic entropy $S$ of any isolated system is a function of the energy $E$ and additional control parameters $Z$ (external magnetic fields, etc.). Thus,  for any isolated system, which by definition has constant $E$ and $Z$, the entropy $S(E,Z)$ is constant, implying that the formulation of the second law given by \Sch{} is wrong.  Furthermore, it is not difficult to verify that the Gibbs entropy satisfies Planck's~\cite{PlanckBook1903} second law for arbitrary continuous DoS, whereas the Boltzmann entropy does not (see Sec.~V in Ref.~\cite{2014HiHaDu} for a detailed discussion).

The \lq Gedankenexperiment\rq~narrated by \Sch{} merely describes a situation in which an otherwise isolated system is allowed to interact temporarily with parts of its environment, whereby the system loses energy to the environment and lowers both its internal energy and its  entropy, as correctly predicted by the Gibbs entropy. This process is perfectly consistent with the second law, since the transfer of energy to the environment is accompanied by an entropy transfer (and possibly entropy creation) that increases the entropy of the environment accordingly. 

\vspace{0.2cm}
\par
(v)   \Sch{} state that \textit{\lq\lq $\TempG$ violates that heat should always flow from the hotter (smaller $1/T$) into the colder (larger
$1/T$) system.\rq\rq{}}  This erroneous heat flow argument repeats similarly incorrect objections in Refs.~\cite{2014VilarRubi,2014FrenkelWarren}. It is often naively assumed that temperature tells us in which direction heat will flow when two initially isolated bodies are placed in thermal contact. This heuristic rule-of-thumb works well in the case of \lq{}normal\rq{} systems that possess a monotonically increasing DoS $\DoS$. But it is not difficult to show that, in general, neither the Gibbs temperature nor the Boltzmann temperature nor any of the other suggested alternatives are capable of specifying uniquely the direction of heat flow when two isolated systems become coupled. One obvious reason is that the microcanonical temperature is not one of the primary thermodynamic state variables of an isolated and,  therefore, does~\emph{not} always uniquely characterize the state of an isolated system before it is coupled to another. To illustrate this explicitly, consider as a simple generic example a system with integrated DoS
\be\label{eq:wiggly_ddos}
\IDoS(E)=\exp \left[\frac{E}{2 \epsilon }-\frac{1}{4} \sin \left(\frac{2E}{\epsilon }\right)\right]+\frac{2 E}{\epsilon },
\ee
where $\eps$ is some energy scale. The associated DoS is non-negative and non-monotonic, $\DoS\equiv \p\IDoS/\p E\ge 0$ for all $E\ge 0$. 
As evident from Fig.~1 in Ref.~\cite{2014HiHaDu}, neither Gibbs nor Boltzmann temperature provide a unique thermodynamic characterization in this case, as the same temperature value $\TempG$ or $\TempB$ can correspond to vastly different energy values. When coupling such a system to a second system, the direction of heat flow can be different for different initial energies of the first system even if the corresponding initial temperatures of the first system may be the equal. It is not difficult to see that qualitatively similar results are obtained for all continuous functions $\DoS\ge 0$ that exhibit at least one local maximum and one local minimum on $(E,\infty)$. This ambiguity reflects the fact that the essential control parameter (thermodynamic state variable) of an isolated system is the energy $E$ and \emph{not} the temperature -- this well-known fact~\cite{Callen} appears to have been overlooked in Ref.~\cite{2014Schneider_Comment,2014VilarRubi,2014FrenkelWarren}.
\vspace{0.2cm}
\par
More generally, this means that microcanonical temperatures do not specify the heat flow between two initially isolated systems and, therefore, naive temperature-based heat-flow arguments~\cite{2014FrenkelWarren,2014VilarRubi,2014Schneider_Comment} cannot be used to judge the Gibbs entropy or the Boltzmann entropy or any other entropy. To identify the most appropriate candidate, one must analyze whether or not the different definitions respect the laws of thermodynamics. Such an analysis is presented in Ref.~\cite{2014HiHaDu} and confirms the conclusions in our paper~\cite{2014DuHi_NatPhys}.

\vspace{0.2cm}
\par
(vi) \Sch{} claim  \textit{\lq\lq $\TempG$ is inconsistent with $T$ defined for canonical ensembles.\rq\rq{}} As evident from this statement, \Sch{} implicitly assume an equivalence of microcanonical and canonical ensembles, ignoring the fact  that it is well-established through a number of rigorous results that these two ensembles are, in general, not equivalent as they refer to completely different physical situations (isolated systems with conserved energy \textit{vs.} non-isolated systems with fluctuating energy due to coupling to an infinite bath). The fact that, in a certain energy range,  the reduced density matrix takes an approximately exponential form with some parameter $\beta$ does \emph{not} imply that $1/\beta$ is equal to the thermodynamic temperature of the whole isolated system. The physically and logically correct approach to the thermodynamics of isolated systems is to start from the microcanonical density operator and use an entropy that is consistent with the laws of thermodynamics~\cite{Gibbs,2014HiHaDu}.

\vspace{0.2cm}
\par
\Sch{} close their comment by stating that \textit{\lq\lq negative absolute temperatures are a well-established concept, which is
not only consistent with thermodynamics, but unavoidable for a consistent description of the
thermal equilibrium of inverted populations.\rq\rq{}} The above remarks, combined with the exact results in~\cite{Gibbs,Khinchin,2014HiHaDu} show, that this statement is incorrect as it is based on several logical errors and false assumptions.\footnote{\Sch{} not only fail to show that negative absolute temperatures are \emph{possible} within a thermodynamically consistent microcanonical framework. They also do not explain why a description based on negative absolute temperatures, if it were possible, should then be \emph{unavoidable}.}

\vspace{0.2cm}
\par
In conclusion, we appreciate constructive criticism and scientific dispute as long as arguments are presented in a sensible manner. We find it questionable when historical facts are misrepresented to discredit opposing opinions. On page 2 of their Comment, \Sch{} wrote \textit{\lq\lq Dunkel and Hilbert in [1] ... choose a non-standard definition of the entropy of microcanonical ensembles using the so-called Hertz entropy (unfortunately named Gibbs entropy in Ref. [1])\rq\rq}. The last part of this sentence suggests that \Sch{} question the integrity of the terminology in our paper~\cite{2014DuHi_NatPhys}.  Assuming that \Sch{} are familiar with the context of Gibbs' book~\cite{Gibbs} and Hertz' work~\cite{1910Hertz_1,1910Hertz_2} through the references provided in our paper~\cite{2014DuHi_NatPhys}, we are surprised that \Sch{} missed the fact that Hertz explicitly acknowledges Gibbs' work which appeared about 10 years earlier, see second paragraph on page 226 in Ref.~\cite{1910Hertz_1} and footnote 2 in Eq. (178)  in Ref.~\cite{1910Hertz_2}. In view of the historical facts,  we believe that it is correct to credit Gibbs, even if previous papers (including two of our own) also used the term \lq{}Hertz entropy\rq{}.


\end{document}